\documentclass[fleqn,10pt]{wlscirep}
\usepackage[utf8]{inputenc}
\usepackage[T1]{fontenc}
\usepackage{comment}
\title{Gouy phase and quantum interference with cross-Wigner functions for matter-waves}

\author[1,2,*]{Lucas S. Marinho}
\author[3]{Pedro R. Dieguez}
\author[4]{Carlos H. S. Vieira}
\author[1]{Irismar G. da Paz}

\affil[1]{Departamento de F\'isica, Universidade Federal do Piau\'i, Campus Ministro Petr\^onio Portela, 64049-550, Teresina, PI, Brazil}

\affil[2]{Departamento de F\'isica, Universidade Federal de Pernambuco, 50670-901 Recife, Pernambuco, Brazil}

\affil[3]{International Centre for Theory of Quantum Technologies, University of Gda\'nsk, Jana Bazynskiego 8, 80-309 Gda\'nsk, Poland}

\affil[4]{Centro de Ci\^{e}ncias Naturais e Humanas, Universidade Federal do ABC,
Avenida dos Estados 5001, 09210-580 Santo Andr\'e, S\~{a}o Paulo, Brazil}

\affil[*]{lucas.marinho@ufpi.edu.br}

\begin{abstract}
The Gouy phase is essential for accurately describing various wave phenomena, ranging from classical electromagnetic waves to matter waves and quantum optics. In this work, we employ phase-space methods based on the cross-Wigner transformation to analyze spatial and temporal interference in the evolution of matter waves characterized initially by a correlated Gaussian wave packet. 
First, we consider the cross-Wigner of the initial wave function with its free evolution, and second for the evolution through a double-slit arrangement. Different from the wave function which acquires a global Gouy phase, we find that the cross-Wigner acquires a Gouy phase difference due to different evolution times. The results suggest that temporal like-Gouy phase difference is important for an accurate description of temporal interference.
Furthermore, we propose a technique based on the Wigner function to reconstruct the cross-Wigner from the spatial intensity interference term in a double-slit experiment with matter waves.   
\end{abstract}
\begin{document}

\flushbottom
\maketitle
%
%
\thispagestyle{empty}

\section*{Introduction}
\label{sec:introduction}

  Gouy phase is a wave phenomenon that appears when waves are confined transversely to their propagation direction, such as in the case of lens focusing and diffraction through slits. It was first observed for classical light waves~\cite{gouy1,gouy2} and it has been recently investigated for the transverse confinement of matter waves~\cite{IGdaPaz_2007,Paz1,Paz2,Hansen}. The Gouy phase and its properties have been extensively studied~\cite{pang2011phase,pang2011gouy,pang2013manifestation,pang2014wavefront}, and the acquired phase is known to depend on the type of transversal confinement and the geometry of the waves. For instance, line-focusing a cylindrical wave propagating from $-\infty$
to $+\infty$ yields a Gouy phase of $\pi/2$, while point-focusing a spherical wave in the same interval yields a Gouy phase of $\pi$~\cite{feng2001}. Gaussian matter wave packets diffracting through small apertures generate a Gouy phase of $\pi/4$~\cite{Paz4}.
Experiments were performed with acoustic and water waves~\cite{Holme2002,Chauvat,KoboldAIphys2023}, surface plasmon-polaritons with non-Gaussian spatial properties~\cite{zhu}, focused cylindrical phonon-polariton wave packets in LiTaO$_3$ crystals, and for Bose-Einstein condensates and electron waves~\cite{schultz2014raman,guzzinati2013observation,schattschneider2014imaging,petersen2013measurement}. Interestingly, the effects of the quantum Gouy phase was also verified with the evolution of two-photon states~\cite{Fickler2022Naturephot}.

Applications, such as in evaluating the resonant frequencies in laser cavities~\cite{siegman}, in phase-matching in strong-field and high-order harmonic generation~\cite{balcou1993phase,lewenstein1995phase,lindner2003high}, and in describing the spatial
profile of laser pulses with high repetition rate~\cite{Lindner} was shown to be feasible. In
addition, an extra Gouy phase appears in optical and matter waves
depending on the orbital angular momentum's magnitude~\cite{allen1992orbital,allen1999iv,guzzinati2013observation,schattschneider2014imaging}. In recent work, it was found that the Gouy
phase may cause nonlocal effects that modify the symmetries of self-organization in atomic systems~\cite{Guo}. This phase may also be useful in communication and optical tweezers using structured light~\cite{Khoury}.
 Gouy phases in matter waves also display potential applications and
can be used in mode converters in quantum information systems~\cite{Zeilinger_Gouy2018PRL}, in the generation of singular electron optics
\cite{petersen2013measurement} and in the study of non-classical (exotic or looped) paths in interference experiments~\cite{Paz3,Vieira_MPLA}. 


Since the Gouy phase of matter waves is directly related to position-momentum correlations~\cite{Paz1,lustosa2020irrealism}, we propose in this work to investigate the role of the Gouy phase employing initial position-momentum correlations by using the cross-Wigner transform to analyze the temporal evolution of matter waves. The cross-Wigner transform has been employed in different contexts such as signal processing~\cite{cohen1995time,cohen2013leid1}, and quantum measurement theory~\cite{de2012weak,de2012reconstruction}. 
In signal processing, for instance, it is used to analyze the time-frequency content of signals as it allows one to analyze signals in both the time and frequency domains simultaneously~\cite{cohen1995time,cohen2013leid1}. 
In quantum theory, the cross-Wigner was employed as an analog of the two-state vector formalism~\cite{aharonov1964time} for continuous variables~\cite{de2012reconstruction} and to derive the weak values of an observable from a complex quasi-probability distribution associated with it~\cite{de2012weak}. Its physical interpretation showed to be that of an interference term in the Wigner distribution of the sum of two different wave functions. Then, the cross-Wigner transform is a suitable formalism to work with spatial and temporal interference of quantum states.

 Temporal interference emerges as an alternative approach to scrutinize entanglement across different degrees of freedom, such as time-energy entanglement. For instance, the pioneering Franson interferometry~\cite{FransonPRL1989} introduces a novel experimental test for local hidden-variable theories centered on time interference, and visibility greater than 70.7\% for this setup is known to indicate a violation of a Bell-type inequality, and recent experimental work has reached $96 \pm 1\%$ without background subtraction for entanglement photon pairs generated by spontaneous parametric down-conversion~\cite{ChenPRL2021}. 
The temporal analog of a double-slit experiment for light waves has also been observed~\cite{Tirole2023NatPhys}, showing a clear signature of spectral oscillations for time-diffracted light and an inversely proportional relation between slit separation and period of oscillations. For experiments with matter wave packets, the same type of effect was observed, demonstrating a one-by-one detection scheme, allowing us to visualize the buildup of the quantum interference pattern of single-photoelectrons which stochastically
arrive at the detector plane~\cite{Kaneyasu2023SciRep}. 

The cross-Wigner transform is used here to  assess the temporal interference in two different scenarios (with and without double-slit arrangements) to understand the role of the Gouy phase in temporal interference. The manuscript is organized as follows. First, we review the cross-Wigner formalism. Next, we introduce our contribution by constructing the cross-Wigner transformation between an initial Gaussian wave packet and its corresponding free-evolved state. We introduce a double-slit setup to understand how spatial interference coming from the diffraction in the slits relates to the temporal interference captured by the Gouy phase through the cross-Wigner distribution. Finally, we propose 
an approach to reconstruct the cross-Wigner in a double-slit experiment from the intensity interference term. In the end 
we present our conclusions.

\section*{Cross-Wigner function}\label{sec:CW properties}

In this section, we review the cross-Wigner transform~\cite{de2012weak}, a generalization of the Wigner function for pre and post-selected ensembles. 
In the quantum phase space, the Wigner distribution
\begin{equation}\label{eq:wstate}
    W(x,k)=\frac{1}{2\pi}\int dy\,e^{-iky}\psi^*(x+y/2)\psi(x-y/2), 
\end{equation}
describes the state $\psi(x)$ of a given system, and it is normalized over all phase space. Also, it provides the marginal probability distribution for momentum $|\phi(k)|^{2} = \int dx\,W(x,k)$, and position $|\psi(x)|^{2} = \int dk\,W(x,k)$ of a system.

The cross-Wigner transform of two functions can be defined as
\begin{equation}
    \mathcal{CW}_{\psi,\phi}(x,k) \equiv \frac{1}{2\pi}\int dy\,e^{-iyk}\phi^{\ast}(x+y/2)\psi(x-y/2),
    \label{eq:def-cw}
\end{equation}
where $\phi$ and $\psi$ are wave functions. 
Interestingly, the appearance of interference terms is described by the cross-Wigner transform
\begin{equation}
     W_{\psi+\phi}(q,p)=W_{\psi}+W_{\phi}+2 \text{Re}[\mathcal{CW}_{\psi,\phi}],
\end{equation} 
where $W_{\psi}$ ($W_{\phi}$) is the Wigner function of state $\psi$ ($\phi$), respectively, and $\text{Re}[\mathcal{CW}_{\psi,\phi}]$ is the real part of cross-Wigner function between $\psi$ and $\phi$, represents an interference term in the Wigner function of the superposition $W_{\psi+\phi}$.
 Also, note that the cross-Wigner
transform $\mathcal{CW}_{\psi,\phi}(x,k)$ reduces to the familiar Wigner distribution when $\phi=\psi$.

The cross-Wigner transform satisfies the following properties
\begin{equation}
    \int   \mathcal{CW}_{\psi,\phi}(x,k)dk = \phi^{\ast}(x)\psi(x) \quad \text{and} \quad  \int  \mathcal{CW}_{\psi,\phi}(x,k)dx = F\phi^{\ast}(k)F\psi(k), \quad \text{where} \quad F\psi(k)=\frac{1}{\sqrt{2\pi}}\int dx\,e^{-ikx}\psi(x)
\end{equation}
is the Fourier transform of $\psi$.
Also, one can check that
\begin{equation}
    \iint  \mathcal{CW}_{\psi,\phi}(x,k) dx dk= \langle\phi|\psi\rangle,
\end{equation}
this is, the cross-Wigner function is not a quasiprobability distribution, for that, one might consider the following complex quasi-probability distribution
\begin{equation}
    \rho_{\phi,\psi}(x,k)\equiv\frac{\mathcal{CW}_{\psi,\phi}(x,k)}{\langle\phi|\psi\rangle},
\end{equation} 
Note that, it is symmetric under permutation of $(\phi,\psi)$
since $\rho_{\phi,\psi}^{\dagger}=\rho_{\psi,\phi}$. Also, it holds that
\begin{equation}
    \iint dx dk \, \text{Re}[\rho_{\psi,\phi}(x,k)] = 1
\end{equation}
and
\begin{equation}
    \iint dx dk \, \text{Im}[\rho_{\psi,\phi}(x,k)] = 0.
\end{equation}
Interestingly, this complex quasiprobability can be connected with the weak values of some observable as long as $|\phi\rangle$ and $|\psi\rangle$ are not orthogonal states. We refer the reader to Ref.~\cite{de2012weak} for a complete discussion on that topic.

 In the following, we explore the cross-Wigner function under free evolution to observe how the Gouy phase appears in that context.

\section*{Cross-Wigner function and Gouy phase in free evolution}
\label{sec: free evolution}

In this section, we show that while the wave function acquires a global Gouy phase, the cross-Wigner acquires a relative Gouy phase if we consider the transformation between the initial and a freely evolved state.

In turn, we consider as the initial state the following correlated Gaussian state of transverse width $\sigma_{0}$
\begin{equation}
\psi_0(x_i)= \frac{1}{\sqrt{\sigma \sqrt{\pi}}}  \exp
\left[-\frac{{x^2_i}}{{2\sigma^2_0}} + \frac{i \gamma
x^2_i}{2\sigma^2_0}\right], \label{psi_0}
\end{equation}
that represents a position-momentum correlated Gaussian state. The initial correlation will be represented by the real parameter $\gamma$ which can take values in the interval $-\infty<\gamma<\infty$
\cite{dodonov2002nonclassical,dodonov2014transmission}. The parameter $\gamma$ ensures that the initial state is
correlated. We obtain for the initial state $\psi_0(x_i)$
that the uncertainty in position is
$\sigma_{xx}=\sigma_{0}/\sqrt{2}$, whereas the uncertainty in
momentum is
$\sigma_{pp}=(\sqrt{1+\gamma^{2}})\hbar/\sqrt{2}\sigma_{0}$ and the
$\sigma_{xp}$ correlations is $\sigma_{xp}=\hbar \gamma/2$. For
$\gamma=0$ we have a simple uncorrelated Gaussian wavepacket with
$\sigma_{xp}=0$. (See methods subsection Position-momentum correlations).

The wave function at $t$ is given
by~\cite{Carol}
\begin{eqnarray}
\psi(x,t)&=&\int_{-\infty}^{\infty} dx_i G(x,t;x_i,0)\psi_0(x_i) , 
\end{eqnarray}
 where
\begin{equation}
G(x,t;x_i,0)= \sqrt{\frac{m}{2\pi i \hbar t}} \exp
\left[\frac{im(x-x_i)^2}{2 \hbar t} \right].  \label{M22}
\end{equation}
The kernel $G(x,t;x_{i},0)$ is the free nonrelativistic propagator for a particle of mass $m$.
After some algebraic manipulation, we obtain 
\begin{gather}\label{psi_free}
\psi(x,t)=\frac{1}{\sqrt{b(t) \sqrt{\pi}}}  \exp
\left(-\frac{x^2}{2 b(t)^2} \right)\exp \left(\frac{i m x^2}{2 \hbar r(t)} + i \mu (t) \right) 
\end{gather}
where
\begin{equation}
b(t)= \frac{\sigma_0}{\tau_0} \left[{ t^2 + \tau^2_0 + 2 t \tau_0
\gamma + t^2 \gamma^2 } \right]^\frac{1}{2}, \label{M8} \quad r(t)= \frac{ \left(  t^2 + \tau^2_0 + 2 t \tau_0 \gamma + t^2 \gamma^2
\right) } { \left[  t \left( 1 +  \gamma^2 \right) + \gamma \tau_0
\right] },  \quad \text{and} \quad \mu(t)= - \frac{1}{2} \arctan \left( \frac{ t } {\tau_0+ \gamma t } \right).
\end{equation}

\begin{figure}[htbp]
\centering
\includegraphics[width=18 cm,height = 5.0 cm]{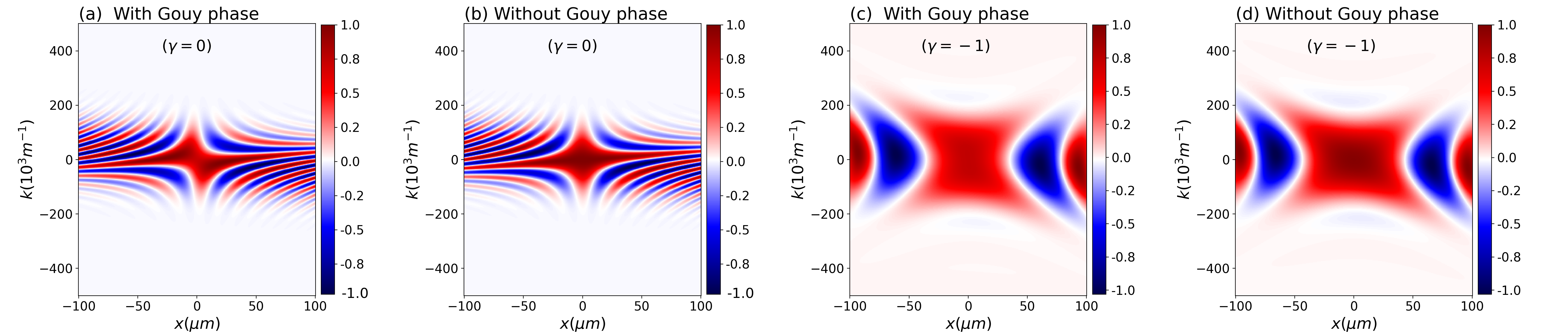}
\caption{Real part of the normalized cross-Wigner ($\text{Re}[\mathcal{CW}]_{\psi,\psi_0}$) as a function of $x$ and $k$ for $t=50$ ms and $\gamma=0$ in (a) and (b), while for (c) and (d) we used $\gamma=-1$. In (a) and (c), we consider the Gouy phase difference, but in (b) and (d), we do not consider it. From these plots, it is evident the importance of the Gouy phase in this cross-Wigner formalism for an accurate and complete description of the quantum state. We have similar results for the imaginary component.}\label{Figure2}
\end{figure}

Here, $b(t)$ is the beam width,
$r(t)$ is the radius of curvature of the wave fronts and $\mu(t)$ is the Gouy phase for the free propagation. The parameter $\tau_{0}=m\sigma_{0}^{2}/\hbar$ is one intrinsic time scale which
essentially corresponds to the time at which the distance of the
order of the wave packet extension is traversed with a speed
corresponding to the dispersion in velocity. It is viewed as a characteristic time for the ``aging" of the initial state
\cite{Carol,solano} since it is a time from which the evolved state
acquires distinguishable properties from the initial state.

Now, we calculate the cross-Wigner between the initial state and the state at $t$, as follows 
\begin{gather}
\mathcal{CW}_{\psi,\psi_0}(x,k) = \frac{1}{2 \pi} \int_{-\infty}^{\infty} d y  e^{-i k y}
\psi^* \left(x+\frac{y}{2}\right) \psi_0\left( x - \frac{y}{2}
\right). \nonumber\\
\end{gather}
After some algebraic manipulation we obtain for its real and imaginary parts, respectively, the following results
\begin{gather}\label{Cw_free_evolution}
 \mathcal{CW}_{\psi,\psi_0}(x,k)=Ne^{-a_{1}x^{2}}e^{-a_{3}k^{2}}e^{a_{5}kx} e^{i\phi},  
\end{gather}
where
\begin{equation}
\phi=a_{2}x^{2}+a_{4}k^{2}+a_{6}kx+\Delta\mu, \quad \text{and} \quad \Delta\mu=\xi(t)-\mu(t) \quad \text{with} \quad  \xi(t)= - \frac{1}{2} \arctan \left( \frac{ \frac{m}{2\hbar r}-\frac{\gamma}{2\sigma_{0}^{2}}} {\frac{1}{2b^{2}}+\frac{1}{2\sigma_{0}^{2}}} \right). 
\end{equation}
Here, $\Delta\mu$ is the cross-Wigner Gouy phase difference for the free evolution. The parameters $N$ and $a_i$ ($i=1,...,6$) are given in methods subsection (Parameters of the cross-Wigner for free-evolution). 

In the following, we consider a wavefunction for neutrons with $m=1.67\times10^{-27}\;\mathrm{kg}$, $\sigma_{0}=7.8\;\mathrm{\mu m}$~\cite{zeilinger1988single}.
In Fig. \ref{Figure2}, we show the real part of the normalized cross-Wigner from eq.(\ref{Cw_free_evolution}) as a function of $x$ and $k$ for $t=50$ ms and two different values of $\gamma$. In Fig. \ref{Figure2}(a) and (b), we consider $\gamma=0$, while for Fig. \ref{Figure2}(c) and (d), we used $\gamma=-1$ . In Fig. \ref{Figure2}(a) and (c), we consider the Gouy phase difference, but in Fig. \ref{Figure2}(b) and (d), we do not consider it. At this point, the importance of the Gouy phase in this cross-Wigner formalism for an accurate and complete description of the quantum state becomes evident. Here, the Gouy phase difference results from the fact that the cross-Wigner is calculated for wave functions at different times, i.e., $\psi(x,t)$ at the time $t>0$ and $\psi_0$ at the time $t=0$. 

For a comparison between the free evolution and evolution through a double slit (see section Cross-Wigner function and Gouy phase in the double-slit experiment), we show, in Fig. \ref{Figure3} (a), the normalized cross-Wigner from eq.(\ref{Cw_free_evolution}) between the initial state and the state at the time $t$ (the free evolution case) as a function of $x$ and $t$ for $k=0$ and $\gamma=0$. We can see a cross-Wigner's peak located around $(x=0,t=0)$, indicating that this is the region where there is greater overlap between $\psi(x,t)$ and $\psi_0(x)$. Nevertheless, the overlapping decreases at high $t$ values, which reduces the cross-Wigner function. 

\section*{Cross-Wigner function and Gouy phase in the double-slit experiment}
\label{sec:double_slit}

%

\begin{figure}[htbp]
\centering
\includegraphics[width=8.5 cm]{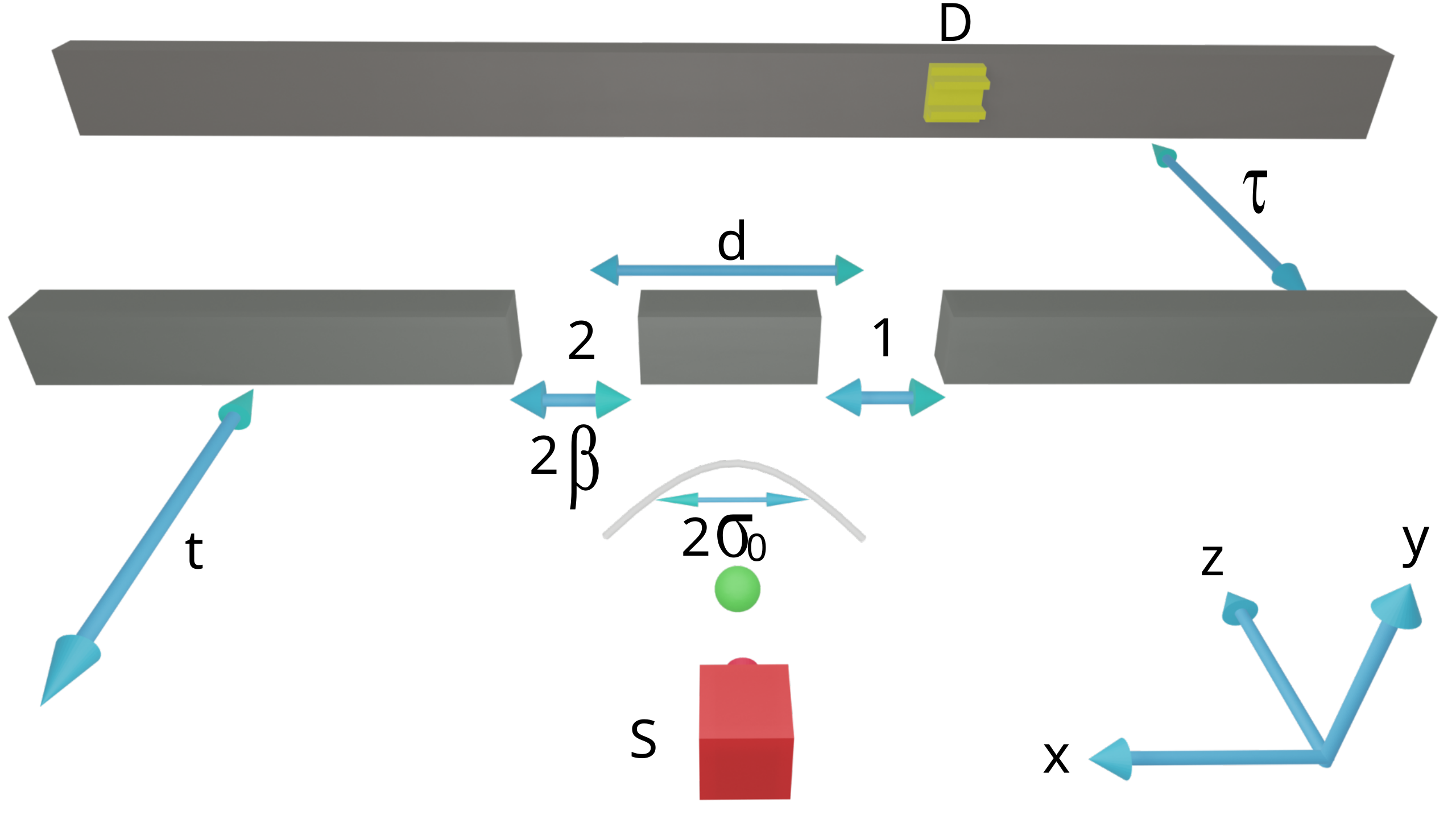}
\caption{Sketch of the double-slit experiment. The source $S$
produces a correlated Gaussian wavepacket with the transverse width
$\sigma_{0}$. The wavepacket propagates during a time $t$ before
attaining the double-slit and during a time $\tau$ from the
double-slit to the detector $D$ in the screen of detection. The slit
transmission functions are taken to be Gaussian of width $\beta$ and
separated by a distance $d$.}\label{Figure4}
\end{figure}

Consider a classical double-slit experiment with an initially
correlated Gaussian wavepacket given by eq. \eqref{psi_0}. We use this setup to study how the spatial correlation generated in the double-slit affects the cross-Wigner distribution. Assume that such coherent correlated Gaussian wavepacket is produced in
the source $S$ and propagates during a time $t$ before arriving at a
double-slit which splits it into two Gaussian wavepackets. After
crossing the grid, the wavepackets propagate during a time $\tau$
before arriving at detector $D$ in the detection screen. In this model, we consider wave effects only in
$x$-direction as we can assume that the energy associated with the
momentum of the particles in the $z$-direction is high enough such
that the momentum component $p_{z}$ is sharply defined, i.e.,
$\Delta p_{z}\ll p_{z}$. Then we can consider a classical behavior
in this direction at velocity $v_{z}$, and hence we can write $z=v_{z}t$
\cite{Paz3}. The sketch of this model is presented in Fig. \ref{Figure4}.

The wavefunctions at the right $1(+)$ and left slit $2(-)$ are given
by~\cite{Carol}
\begin{eqnarray}
\psi(x,t,\tau)&=&\int_{-\infty}^{\infty} dx_jG(x,t+\tau;x_j,t)F(x_j\pm d/2)\psi(x_j,t) , 
\end{eqnarray}
 where
\begin{equation}
G(x,t+ \tau;x_j,t)= \sqrt{\frac{m}{2\pi i \hbar \tau}} \exp
\left[\frac{im(x-x_j)^2}{2 \hbar \tau} \right]  , \quad \text{and} \quad  F(x_j \pm d/2)= \frac{1}{\sqrt{\beta \sqrt{\pi}}}  \exp
\left[-\frac{im(x_j \pm d/2)^2}{2 {\beta}^2} \right] .
\end{equation}
$F(x_{j}\pm d/2)$ describes the double-slit transmission functions
which are taken to be Gaussian functions of width $\beta$ separated by a
distance $d$. To obtain analytic expressions for the
wavefunction, Wigner and Cross-Wigner functions in the screen of detection, we use a Gaussian transmission function instead of a top-hat transmission one because both a Gaussian transmission function represents a good approximation to the experimental reality and it is mathematically simpler to treat than a top-hat
transmission function.

The wavefunction that passed through slit $1(+)$ is given by
\begin{eqnarray}
\psi_1(x,t,\tau)&=&\frac{1}{\sqrt{B \sqrt{\pi}}}  \exp
\left[-\frac{{(x+D/2)}^2}{{2 B^2}} \right] \exp \left(\frac{i m x^2}{2 \hbar R} + i \Delta x + i \theta
+ i \mu^{\prime} \right) , 
\end{eqnarray}
where
\begin{equation}
\mu^{\prime}(t,\tau)= - \frac{1}{2} \arctan \left[ \frac{ t + \tau \left( 1 +
\frac{\sigma^2_0}{\beta^2} + \frac{t \hbar \gamma }{ m \beta^2 }
\right) } { \tau_0 \left( 1 - \frac{t \tau \sigma^2_0}{\tau_0
\beta^2} \right) + \gamma \left( t + \tau \right)} \right],
\end{equation}
describes the time-dependent Gouy phase in double-slit. Differently from the
results obtained in Ref.~\cite{solano}, all the parameters above are
affected by the correlation parameter $\gamma$ and can be checked explicitly in the methods subsection (Parameters of the cross-Wigner for evolution through a double-slit).

\begin{figure}[htbp]
\centering
\includegraphics[width=8.5 cm, height = 6.0 cm]{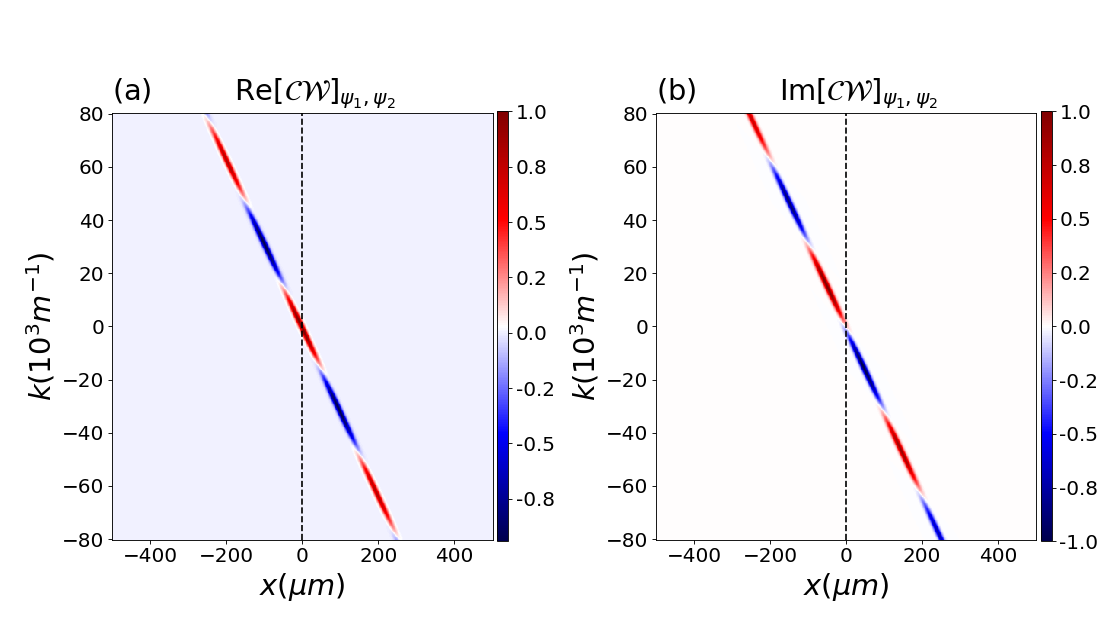}
\caption{Normalized cross-Wigner transform (a) Real part ($\text{Re}[\mathcal{CW}]_{\psi_1,\psi_2}$) as a function of $x$ and $k$ for $t=50$ ms,  for $\tau=50$ ms and $\gamma=0$. The real and imaginary components of the cross-Wigner function only differ by a $\pi/2$ phase. Then, one component can be obtained from the other part through phase displacement in the $(x,k)$ space and (b) the imaginary value ($\text{Im}[\mathcal{CW}]_{\psi_1,\psi_2}$) for the same parameters.}\label{Figure5}
\end{figure}

The parameter $B(t,\tau)$ is the beam
width for the propagation through one slit, $R(t,\tau)$ is the
radius of curvature of the wavefronts for the propagation through
one slit, $D(t,\tau)$ is the separation between the wavepackets
produced in the double-slit. $\Delta(t,\tau)x$ is a phase that
varies linearly with the transverse coordinate. $\theta(t,\tau)$ and
$\mu^{\prime}(t,\tau)$ are the time-dependent phases and they are relevant
only if the slits have different widths. $\mu^{\prime}(t,\tau)$ is the Gouy
phase for the propagation through one slit. Differently from the
results obtained in Ref.~\cite{solano}, all the parameters above are
affected by the correlation parameter $\gamma$. For the left slit $2(-)$, we have just to substitute the parameter
$d$ with $-d$ in the expressions corresponding to the wave passing
through the first slit.

Having obtained the wavefunctions, we calculate the cross-Wigner transform. First, we calculate the cross-Wigner between the states $\psi_1(x,t,\tau)$ and $\psi_2(x,t,\tau)$ at the detection screen. It is given by
\begin{eqnarray}
\mathcal{CW}_{\psi_1,\psi_2}(x,k) = \frac{1}{2 \pi} \int_{-\infty}^{\infty} d y  e^{-i k y}
\psi_1^* \left(  x+\frac{y}{2} \right) \psi_2\left( x - \frac{y}{2}
\right), \nonumber\\
\end{eqnarray}
After some algebraic manipulation, we obtain for the real and imaginary parts the following results

\begin{gather}\label{CW_double_slit}
 \mathcal{CW}_{\psi_1,\psi_2}(x,k)=\frac{2}{\pi}\exp\left[-\left(\frac{x^2}{B^2}+\left(k+\frac{mx}{\hbar
R}\right)^2 B^2\right)\right]\exp \left[i\left(k+\frac{mx}{\hbar R}\right)D-2i\Delta
x\right].  
\end{gather}

The results above are independent of the Gouy phase because both wave functions evolve at the same time concerning the initial state and acquire the same Gouy phase $\mu^{\prime}(t,\tau)$. 

In the following, we consider the neutron parameters $m=1.67\times10^{-27}\;\mathrm{kg}$, $\sigma_{0}=7.8\;\mathrm{\mu m}$, $\beta=7.8\;\mathrm{\mu m}$ and $d=100\;\mathrm{\mu m}$. Fig. \ref{Figure5} shows the plots for the real and imaginary part of the cross-Wigner between the states of propagation through the slits as a function of $x$ and $k$ for $t=50$ ms,  $\tau=50$ ms, and $\gamma=0$. The real and imaginary components of the cross-Wigner function only differ by a $\pi/2$ phase, as we can see from Eq. (\ref{CW_double_slit}). Note that the imaginary part can then be obtained from the real part through phase displacement in the $(x,k)$ space.

\begin{figure}[htbp]
\centering
\includegraphics[width=8.5 cm, height = 6.0 cm]{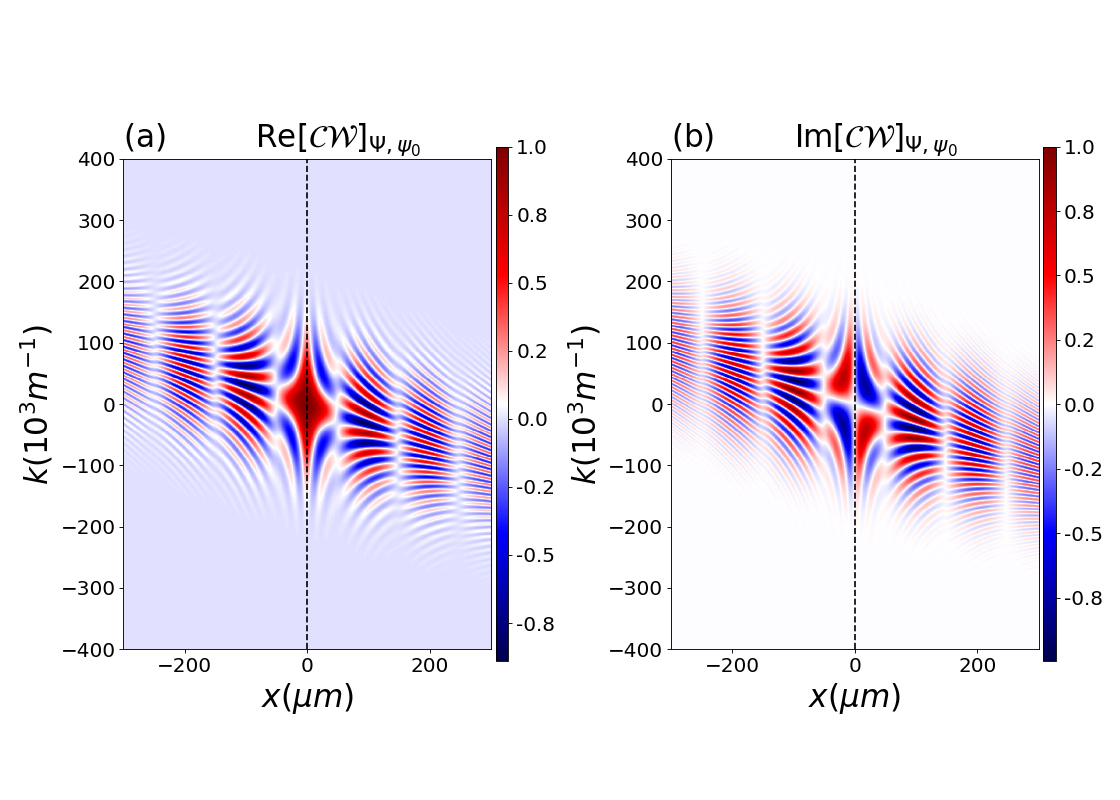}
\caption{Normalized cross-Wigner. (a) The real part ($\text{Re}[\mathcal{CW}]_{\Psi,\psi_0}$) as  a function of $x$ and $k$ for $t=50$ ms,  for $\tau=50$ ms and $\gamma=0$  (b) The imaginary part ($\text{Im}[\mathcal{CW}]_{\Psi,\psi_0}$) with the same parameters .}\label{Figure7}
\end{figure}

Next, we calculate the cross-Wigner between the initial state and the state at the detection screen, which is given by
\begin{gather}
\mathcal{CW}_{\Psi,\psi_0}(x,k) = \frac{1}{2 \pi} \int_{-\infty}^{\infty} d y  e^{-i k y}
\Psi^{*} \left(  x+\frac{y}{2} \right) \psi_0\left( x - \frac{y}{2}
\right), \nonumber \\ 
\end{gather}
where
\begin{equation}
\Psi(x,t,\tau)= \frac{\psi_1 (x,t,\tau) + \psi_2 (x,t,\tau)}
{\sqrt{2 + 2 \exp \left[ -  \frac{D^2}{4 B^2}  - \Delta^2 B^2
\right]}}, \label{superposition_state}
\end{equation}
is the normalized wave function at the screen of detection of the double-slit experiment.
After some algebraic manipulation we obtain  for its real and imaginary parts the following results 
\begin{gather} \mathcal{CW}_{\Psi,\psi_0}=N^{\prime}e^{-\frac{D^{2}}{8B^{2}}}e^{b_1}e^{-b_{2}k^{2}}e^{-b_{3}x^{2}}e^{-b_{4}kx} [e^{b_{5}x} e^{b_6 k}e^{i(\phi_1+\phi_2)}+e^{-b_{5}x} e^{-b_6 k}e^{i(\phi_1-\phi_2)}],    \label{CW_psi0_ps1+ps2}
\end{gather}
and 
where
\begin{equation}
\phi_1=b_{7}x^{2}+b_{8}k x+b_{9}+b_{10}k^{2}-\theta+\Delta\mu^{\prime}, \quad \phi_2=b_{11}x+b_{12}k,
\quad \text{and} \quad  \Delta\mu^{\prime}(t,\tau)=\xi^{\prime}(t,\tau)-\mu^{\prime}(t,\tau)
\end{equation}
with 
\begin{equation}
\xi^{\prime}(t,\tau)= - \frac{1}{2} \arctan \left( \frac{ \frac{m}{2\hbar R}-\frac{\gamma}{2\sigma_{0}^{2}}} {\frac{1}{2B^{2}}+\frac{1}{2\sigma_{0}^{2}}} \right). 
\end{equation}
Here, $\Delta\mu^{\prime}$ is the cross-Wigner Gouy phase difference for the propagation through the slit. The parameters $N^{\prime}$ and $b_i$ ($i=1,...,12$) are given in the methods subsection (Parameters of the cross-Wigner for evolution through a double-slit). 
Here we observe that the result is dependent on the cross-Wigner Gouy phase difference $\Delta\mu^{\prime}(t,\tau)$ because as in the free evolution, the cross-Wigner is calculated for states at different times. 

\begin{figure}[htbp]
\centering
\includegraphics[width=18 cm, height = 5 cm]{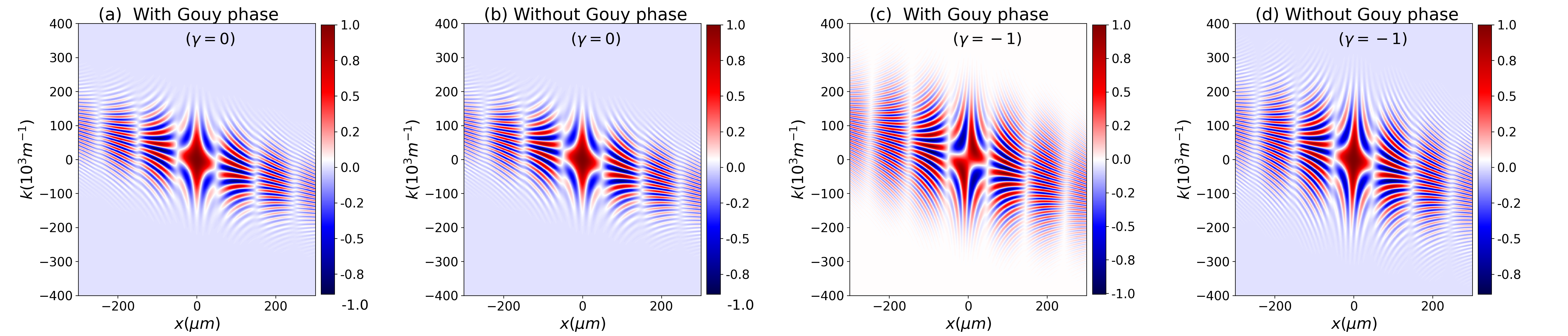}
\caption{Normalized cross-Wigner real part ($\text{Re}[\mathcal{CW}]_{\Psi,\psi_0}$) as a function of $x$ and $k$ for $t=50$ ms, $\tau=50$ ms and $\gamma = 0$ in (a) and
(b), while for (c) and (d) we used $\gamma = -1$. In (a) and (c), we consider the Gouy phase difference, and in (b) and (c), we do not consider it. As for the free propagation context, the Gouy phase is crucial for providing a precise and accurate description of the cross-Wigner. We change the value of $\gamma$ to -1, in this plot, since the phase effect is more noticeable than for the case where $\gamma=0$ (See methods subsection Gouy phase behavior with correlation parameter bellow). We have similar results for the imaginary component.}\label{Figure8}
\end{figure}

\begin{figure}[htbp]
\centering
\includegraphics[width=8.5 cm,height = 6.0 cm]{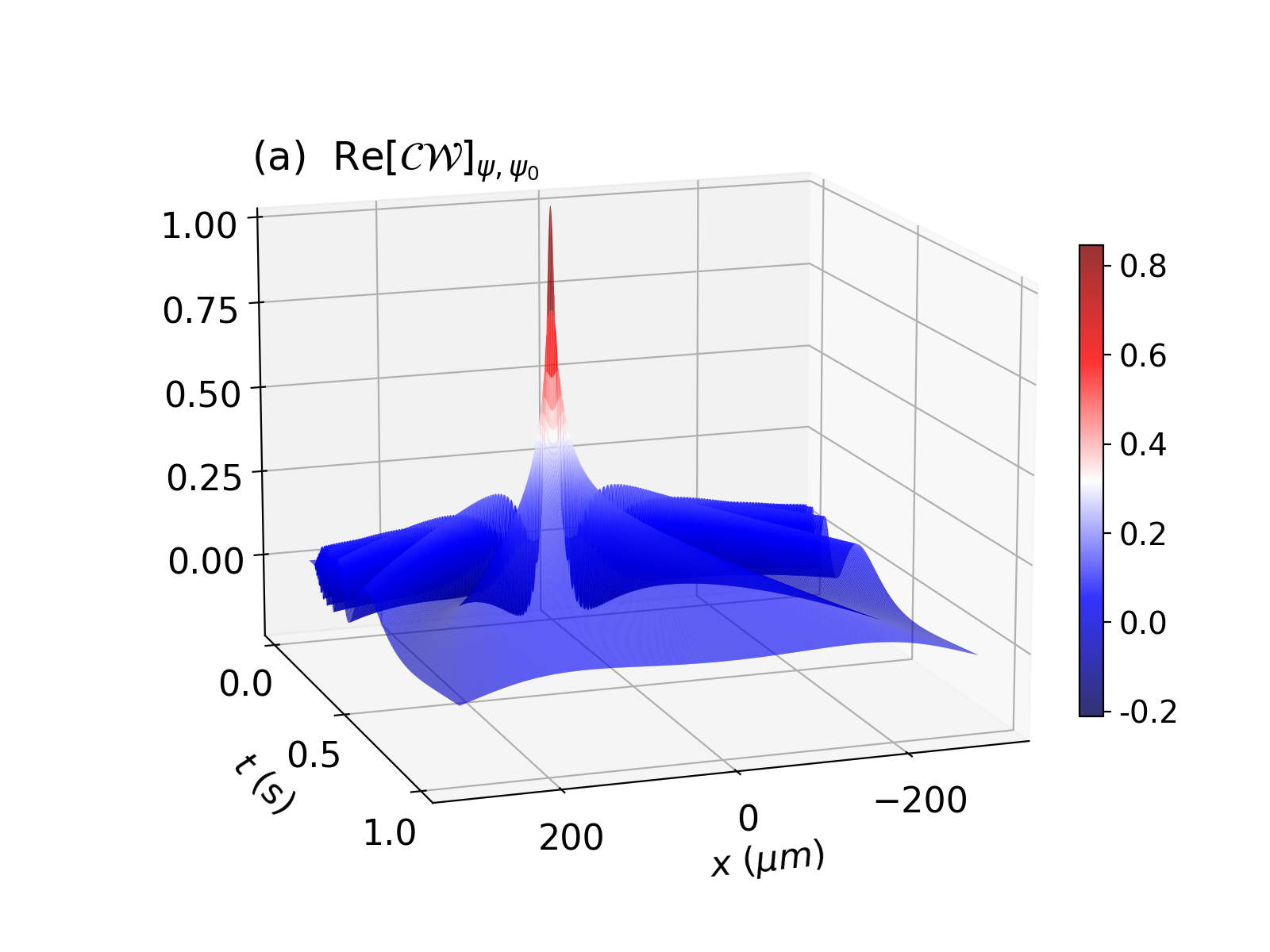}
\includegraphics[width=8.5 cm, height = 6.0 cm]{
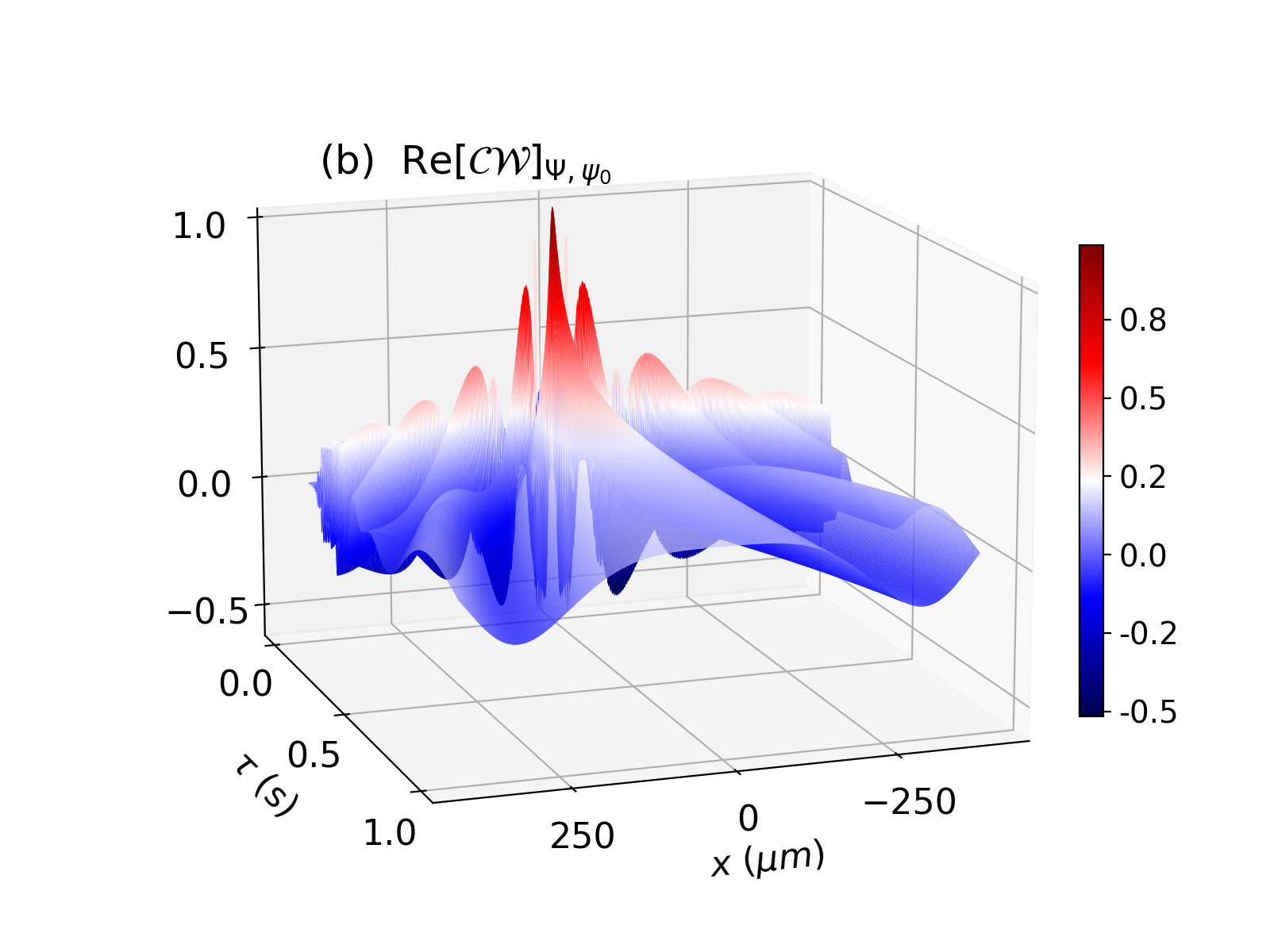}
\caption{ (a) Normalized cross-Wigner real part ($\text{Re}[\mathcal{CW}]_{\psi,\psi_0}$) for the free evolution case as a function of $x$ and $t$ for $k=0$ and $\gamma=0$. Around the $(x=0,t=0)$, exists one maximum indicating that in this region there is greater overlap between $\psi(x,t)$ and $\psi_0(x)$. (b) Normalized cross-Wigner real part ($\text{Re}[\mathcal{CW}]_{\Psi,\psi_0}$) for evolution through the double-slit as a function of $x$ and $\tau$ for $t=50$ ms, $k=0$ and $\gamma=0$. The cross-Wigner real part in this case exhibits more oscillatory behavior than in the free propagation case (a), since the state at the detection screen, in this current scenario, is delocalized, characterized by the superposition of two Gaussian states.
  }\label{Figure3}
\end{figure}

By considering the same parameters of a wave of neutrons used above we show, in Fig. \ref{Figure7}, the cross-Wigner between the initial state and the state at the screen of detection as a function of $x$ and $k$ for $t=50$ ms, $\tau=50$ ms and $\gamma=0$. In (a) we see the normalized cross-Wigner real part and in (b) imaginary part from eq. from eq.(\ref{CW_psi0_ps1+ps2}). Fig. \ref{Figure8} illustrates how the Gouy phase difference affects the cross-Wigner function's real part and highlights how crucial it is for a precise cross-Wigner description. In this plot, we change the value of $\gamma$ to -1 since the effect is more noticeable than for the case where $\gamma=0$ (see appendix Gouy phase behavior with correlation parameter). 
For the imaginary component, our results are similar. 

The normalized cross-Wigner (real part) between the initial state and the state at the screen of detection as a function of $x$ and $\tau$ for $t=50$ ms, $k=0$ and $\gamma=0$ is shown in Fig. \ref{Figure3} (b). Compared with Fig. \ref{Figure3} (a), the cross-Wigner real part, in this situation, has more oscillatory behavior than in the free propagation case, since in this current scenario, the state at the detection screen is delocalized, characterized by the superposition of two Gaussian states in eq. (\ref{superposition_state}) since the non-vanishing region of cross-Wigner depends mainly on the overlapping between the two functions considered in its definition, this extra oscillatory behavior is due to the fact that, in comparison with the free evolution case, in the double-slit setup, the state at the detection screen is more delocalized, which results in non-null overlapping between this state and the initial state over the largest region and consequently an extra oscillatory behavior of a large region.

\subsection*{Intensity and cross-Wigner reconstruction}\label{sec:CW_reconstrution}


Since the cross-Wigner function is not well known by many physicists, in this part, we will show how to reconstruct the cross-Wigner function from an adaptation of the currently used techniques to measure the Wigner function. From this procedure, we highlight how this complex probability distribution can
be obtained experimentally, showing its connection with observed physical quantities for matter waves and indicating that this distribution has a physical content and is not merely one
mathematical function within this formalism. To this end, and for pedagogical reasons, we considered one of the simplest possible scenarios, the double-slit setup.  
%
As mentioned before, integrating $W(x,k)$ over $k$ results in 
\begin{equation}
    I(x) = \int_{-\infty}^{+\infty}W(x,k) dk,
\end{equation}
which correspond to the interference pattern $ I(x)= |\psi_1+\psi_2|^2$ in the double-slit experiment. If we consider a rotated version of the Wigner function~\cite{JanickeJMO1995}, $W_{\theta}(x,k) = W(x\cos\theta-k\sin\theta,x\sin\theta+k\cos\theta) $, where $\theta$ is the angle of rotation in phase-space. This rotated Wigner function results in the following spatial interference pattern 
\begin{equation}\label{Itheta}
    I_{\theta}(x) = \int_{-\infty}^{+\infty}W_{\theta}(x,k) dk.
\end{equation}
Inverting the equation (\ref{Itheta}), the Wigner function can be reconstructed from the continuous set of interference pattern $I_{\theta}(x)$ through~\cite{VogelPRA1989,JanickeJMO1995} 
\begin{gather}\label{inverse_radon}
  W(x,k)   = \frac{1}{4\pi^2} \int_{-\infty}^{+\infty} dx' \int_{-\infty}^{+\infty} |r| dr  \int_{0}^{\pi} 
  d\theta I_{\theta}(x') e^{ir(x'-x\cos\theta-k\sin\theta)} ,
\end{gather}
know as the inverse Radon transformation~\cite{Herman2009}.

\begin{figure}[htbp]
\centering
\includegraphics[width=18 cm, height = 5 cm]{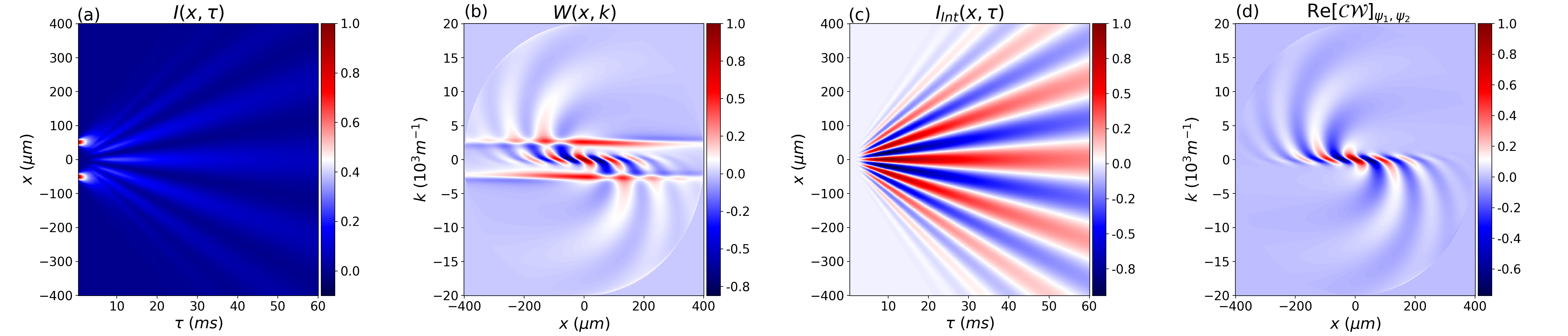} 
\caption{Cross-Wigner reconstruction procedure. In (a) we represent the interference pattern $I(x,\tau)$ between $\psi_1$ and $\psi_2$ in the double-slit experiment for various propagation times $\tau$. (b) Through $I(x,\tau)$ we reconstruct the Wigner function $W(x,k)$.
 (c) On the other hand, from the interference term $I_{\text{Int}}(x,\tau)$ we obtain the real part of the cross-Wigner ($\text{Re}[\mathcal{CW}]_{\psi_1,\psi_2}$)  in (d).}\label{Figure10}
\end{figure}

Generally, in optics experiments the balanced homodyne detection technique~\cite{LvovskyPRL2001} is used to access the rotated basis ($x' =x\cos\theta+k\sin\theta;k'=-x\sin\theta+k\cos\theta$) to obtain the continuous set of interference pattern $I_{\theta}(x)$. On the other hand, another mechanism for mixing the $x$ and $k$ variables is the free evolution of a particle~\cite{KurtsieferNature1997}, with the time propagation playing the role of the rotation angle $\theta$. We emphasize that the free evolution was the technique used as a mixing mechanism between the $x$ and $k$ variables in this work. In Ref.~\cite{KurtsieferNature1997}, the authors reconstruct the Wigner function from the interference pattern $I(x,\tau)$ for various propagation times $\tau$ from the double-slit to the detection screen. We will apply the same kind of approach in our work.

Since the real part of the cross-Wigner transform is one of the interference terms of the standard Wigner distribution of the sum $\psi_1+\psi_2$~\cite{de2012weak}, we can reconstruct the real part of the cross-Wigner function from the interference term $I_{\text{Int}}(x,\tau)$ given by
\begin{gather}
I_{\text{Int}}(x,\tau)=I(x,\tau)-\big[I_1(x,\tau)+I_2(x,\tau)\big] 
= 2\sqrt{I_1(x,\tau)I_2(x,\tau)}\cos(\Phi_{12}),
\end{gather}
where 
\begin{gather}
I_1(x,\tau)=|\psi_1(x,\tau)|^2, \quad I_2(x,\tau)=|\psi_2(x,\tau)|^2 , \quad \text{and} \quad \Phi_{12} = 2\Delta x  
\end{gather}
is the phase-difference between the states $\psi_1$ and $\psi_2$.

In Fig. \ref{Figure10}, we can see how the cross-Wigner reconstruction procedure works. In Fig. \ref{Figure10}(a) we represent the interference pattern $I(x,\tau)$ between $\psi_1$ and $\psi_2$ in the double-slit experiment for various propagation times $\tau$.
As argued in~\cite{KurtsieferNature1997}, only a limited range of evolution times is available in such an experiment. Then,  we should choose an appropriate range of $\tau$, determined by the information on coherence properties of the superposition of the two states $\psi_1$ and $\psi_2$. In Fig. \ref{Figure10}(b), through $I(x,\tau)$ we reconstruct the Wigner function $W(x,k)$, where we can see the characteristic Wigner function of superposition between two Gaussian states, with typical Gaussian-like probabilities located at two different regions of phase-space, and between them an additional interference term that can take negative values. In Fig. \ref{Figure10}(c), on the other hand, from the interference term $I_{\text{Int}}(x,\tau)$ we obtain the real part of the cross-Wigner ($\text{Re}[\mathcal{CW}]_{\psi_1,\psi_2}$) displayed in Fig. \ref{Figure10}(d).


\section*{Conclusions}
\label{sec:conclusion}

We analyzed in this work the role of the Gouy phase difference, for an accurate and complete description of the cross-Wigner transform for matter waves characterized by an initially correlated Gaussian wave packet. In contrast to usual wave functions that only present a global Gouy phase, here we discussed that it is relevant for the cross-Wigner distribution as it becomes a relative phase in this scenario. Moreover, we showed how the Gouy phase is affected by starting the evolution with initial position-momentum correlations that can also amplify the temporal interference between both wave functions. We emphasized that, as the Gouy phase difference is not present in the complex quasi-probability distribution it can incorrectly be neglected in the cross-Wigner function.

We have examined the cross-Wigner between the initial state and the free-evolved one, as well as the evolution through the double-slit arrangement. A temporal Gouy phase difference results from the fact that the cross-Wigner is calculated for wave functions at different times, i.e., $\psi(x,t)$ at the time $t>0$
and $\psi_0$ at the time $t=0$. Then, the cross-Wigner transform provides an interesting tool to asses temporal and spatial interference effects. However, a cross-Wigner is not an absolute indicator to classify whether the type of correlation is spatial and/or temporal, and in general we can have a contribution from both effects. We observe that the phase effect is highlighted when the initial state is contractive because it works as a squeezing agent in this condition, increasing further the transverse confinement of the wavepacket with respect to its propagation direction. We also suggest a method for reconstructing
the cross-Wigner function, fully compatible with current experimental technology~\cite{KurtsieferNature1997}, from the intensity interference term in a double-slit experiment, unveiling then, a relationship between this function and observed physical quantities for matter-waves. 

Based on the results discussed here, one natural extension of this method can be employed to reconstruct cross-Wigner functions with different times of propagation. This idea can be worked on by introducing interferometers like the Franson time interference fringes and temporal beam-splitters~\cite{mendoncca2003temporal} since a sequence of time refraction processes is shown to lead to temporal interference effects. With that, we can then produce interference at the output and use it to reconstruct the cross-Wigner transform. 

It is noteworthy that the Franson interferometer was employed to provide security in a protocol for large-alphabet quantum key distribution~\cite{Howell2007PRL}, where energy-time entanglement is shown to be robust to transmission over large distances in optical fiber. To this end, our work suggests that cross-Wigner formalism possibly can also be used to study temporal properties and provide new insights for these applications.  We let this reconstruction, as well as some possible interesting connections with weak values and cryptography to future investigation.

\section*{Methods}

This section is devoted to discussing in more detail the physical meaning of the position-momentum correlation employed in this work, how it affects the Gouy phase, the basic derivations, and the interpretation of each parameter that appears in the two cross-Wigner functions analyzed here.

\subsection*{Position-momentum correlations}
The correlated Gaussian state from Eq. (\ref{psi_0}) was introduced in Ref.~\cite{DODONOV1980PLA}, where the real parameter $\gamma$ ensures that the initial state is correlated. In this way, for this initial state
the uncertainty in position and momentum is given by $\sigma_{xx}(0)=\sigma_{0}/\sqrt{2}$ and $\sigma_{pp}(0)=(\sqrt{1+\gamma^{2}})\hbar/\sqrt{2}\sigma_{0}$, whereas their covariance becomes
\begin{gather}
    \sigma_{xp}(0)= \langle \psi_0 |( \hat{x}\hat{p} +\hat{p}\hat{x})/2  | \psi_0 \rangle -\langle \psi_0 |\hat{x} |\psi_0 \rangle \langle \psi_0 |\hat{p} |\psi_0 \rangle  =\hbar\gamma/2.
\end{gather}
Exploring the correlation coefficient between $\hat{x}$ and $\hat{p}$, i.e., $r=\sigma_{xp}/\sqrt{\sigma_{x}\sigma_{p}}\,(-1\leq r\leq1)$, the $\gamma$ parameter turns out to be $\gamma=r/\sqrt{1+r^{2}}\,(-\infty\leq \gamma\leq\infty)$, illustrating the physical meaning of the $\gamma$ as a parameter that encoded the initial correlations between $\hat{x}$ and $\hat{p}$ for the initial state. For a particular case where $\gamma=0$, we have a simple initial uncorrelated Gaussian wavepacket. This position-momentum correlation was initially investigated in Ref.~\cite{bohm1951quantum}. Note that in this correlation definition, it was explicitly considered the quantized operators $\hat{x}$ and $\hat{p}$, with $[\hat{x},\hat{p}]=i\hbar$. Also, it was employed the symmetrization procedure to transform the product operator $\hat{x}\hat{p}$ into a hermitian operator $(\hat{x}\hat{p} +\hat{p}\hat{x})/2$. 
From a practice point of view, the origin in this parameter can be seen as due to an atomic beam propagation along a transverse harmonic potential that effectively acts as a thin lens which leads to a quadratic phase shift in the initial state \cite{JanickeJMO1995}. Furthermore, Gaussian correlated packets have been used in many contexts, for example in quantum optics \cite{Campos1999JMO}.

\begin{figure}[htpb]
\centering
\includegraphics[width=18 cm, height = 5 cm]{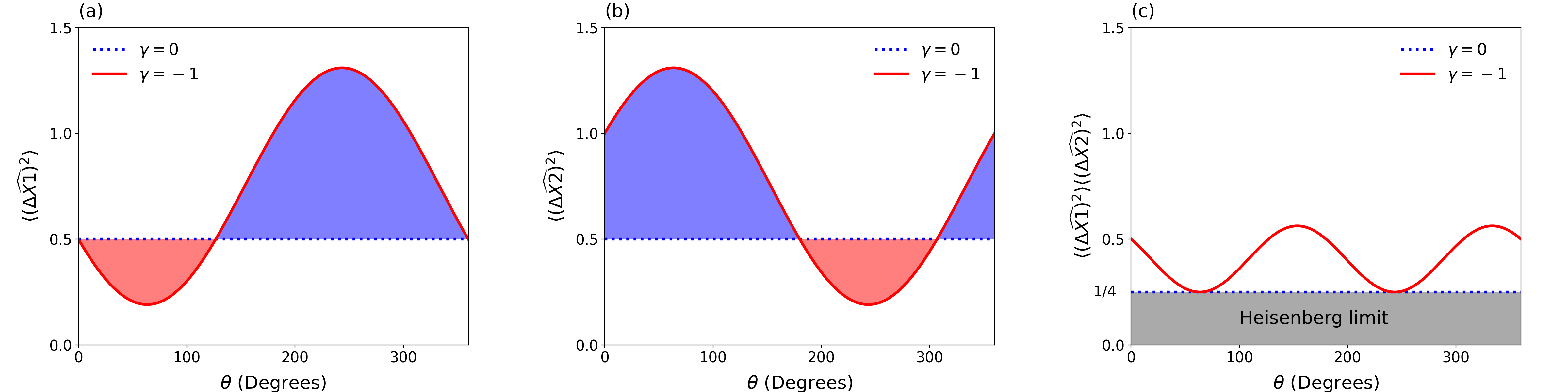} 
\caption{Variances of the operators $\widehat{X1}$ (a) and $\widehat{X2}$ (b) as a function of the rotation angle $\theta$ . For some intervals of $\theta$ the contractive state is squeezed on the $\widehat{X1}$ quadrature  
and spread on the $\widehat{X2}$ quadrature in comparison with the standard Gaussian state.}\label{FigSqueeing}
\end{figure}

In Ref.~\cite{marinho2020squeezing}, it was shown that the
variances in the position and momentum for the dimensionless operators $\hat{x}$ and $\hat{p}$ are given by  
\begin{equation}
    \langle \psi_0 |(\Delta \hat{x})^2  | \psi_0 \rangle = \frac{1}{2};  \quad  \langle \psi_0 |(\Delta \hat{p})^2  | \psi_0 \rangle = \frac{1+\gamma^2}{2},
\end{equation}
which shows that in terms of these conventional operators, the correlated Gaussian state is not squeezed. On the other hand, in terms of the generalized quadratures $\widehat{X1}$ and $\widehat{X2}$, which are defined in terms of  $\hat{x}$ and $\hat{p}$ through a rotation in the phase space by an angle $\theta$ ($\widehat{X1}= \cos\theta \hat{x} + \sin\theta \hat{p}$; $\widehat{X2}= -\sin\theta \hat{x} + \cos\theta \hat{p}$ ), this state presents
squeezing as was shown in Ref.~\cite{marinho2020squeezing}. The variances of the new operators calculated concerning the correlated Gaussian state are
\begin{equation}
    \langle \psi_0 |(\Delta \widehat{X1})^2  | \psi_0 \rangle = \frac{1}{2}[1+\gamma\sin 2\theta + \gamma^2 \sin^2 \theta] \quad  \text{and} \quad \langle \psi_0 |(\Delta \widehat{X2})^2  | \psi_0 \rangle = \frac{1}{2}[1-\gamma\sin 2\theta + \gamma^2 \cos^2 \theta].
\end{equation}

In Fig. \ref{FigSqueeing}, we show, respectively, the variances of the operators $\widehat{X1}$ (a) and $\widehat{X2}$ (b) as a function of the rotation angle $\theta$ for the initially correlated Gaussian state. The solid lines correspond to the variances for $\gamma =-1.0$
(which represents a contractive state) and the dash-dotted lines correspond to the variances for $\gamma  =0$ (the standard Gaussian state). We can observe that for some intervals of $\theta$ the contractive state is squeezed on the $\widehat{X1}$ quadrature  
and spread on the $\widehat{X2}$ quadrature in comparison with the standard Gaussian state. Therefore, it is in this sense that values of $\gamma = -1$  works as a squeezing agent. We also include in (c) the uncertainty relation.

\subsection*{Gouy phase behavior with correlation parameter}\label{GouyphaseAppendix}

In the following, we describe how the time behavior of the Gouy phase is more pronounced for some values of the correlation parameter. In this setting, the Gouy phase is even more important for an accurate description of the cross-Wigner function.
\begin{figure}[htpb]
\centering
\includegraphics[width=8.0 cm,height = 6.0 cm]{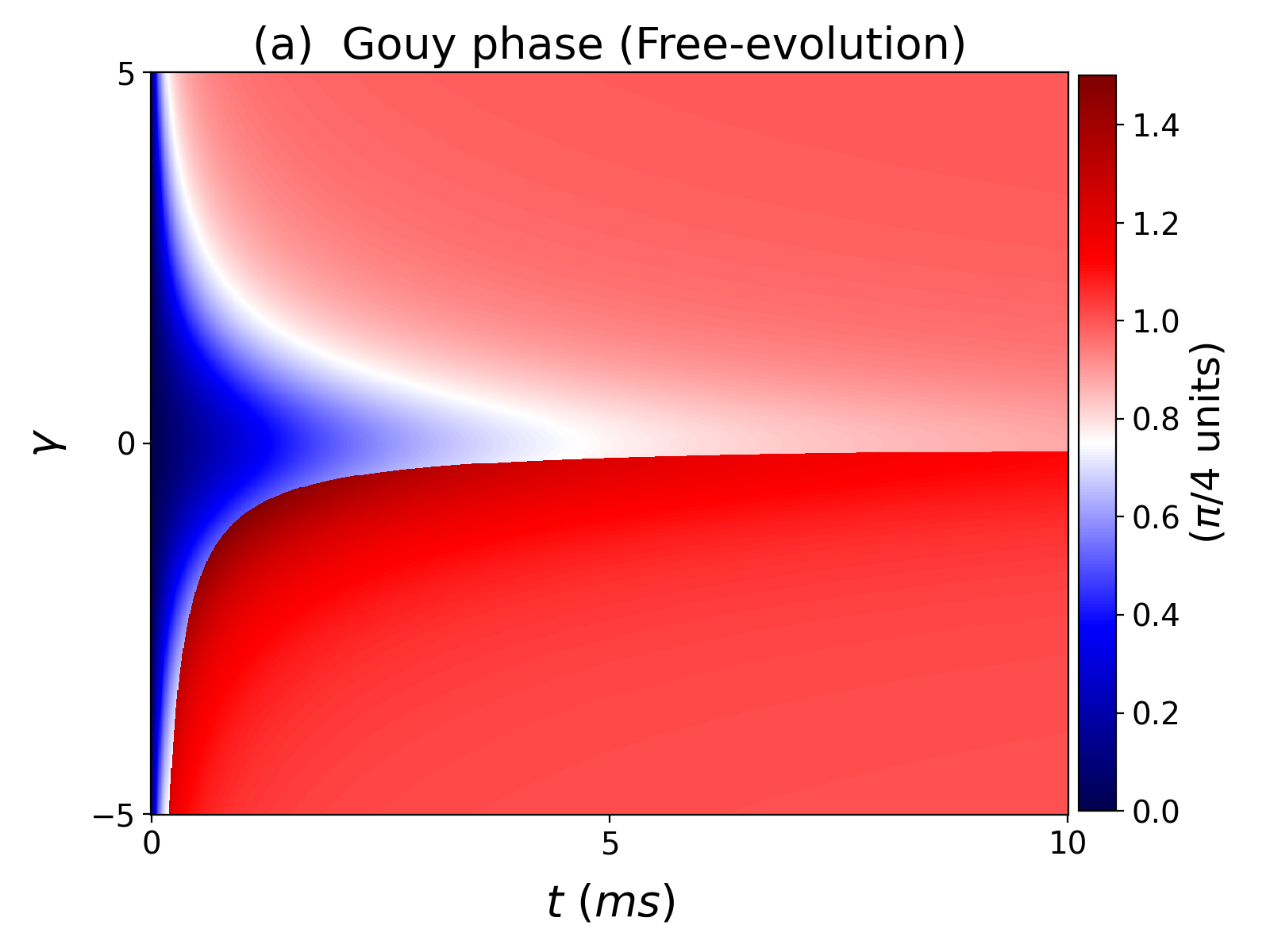}
\includegraphics[width=8.0 cm, height = 6.0 cm]{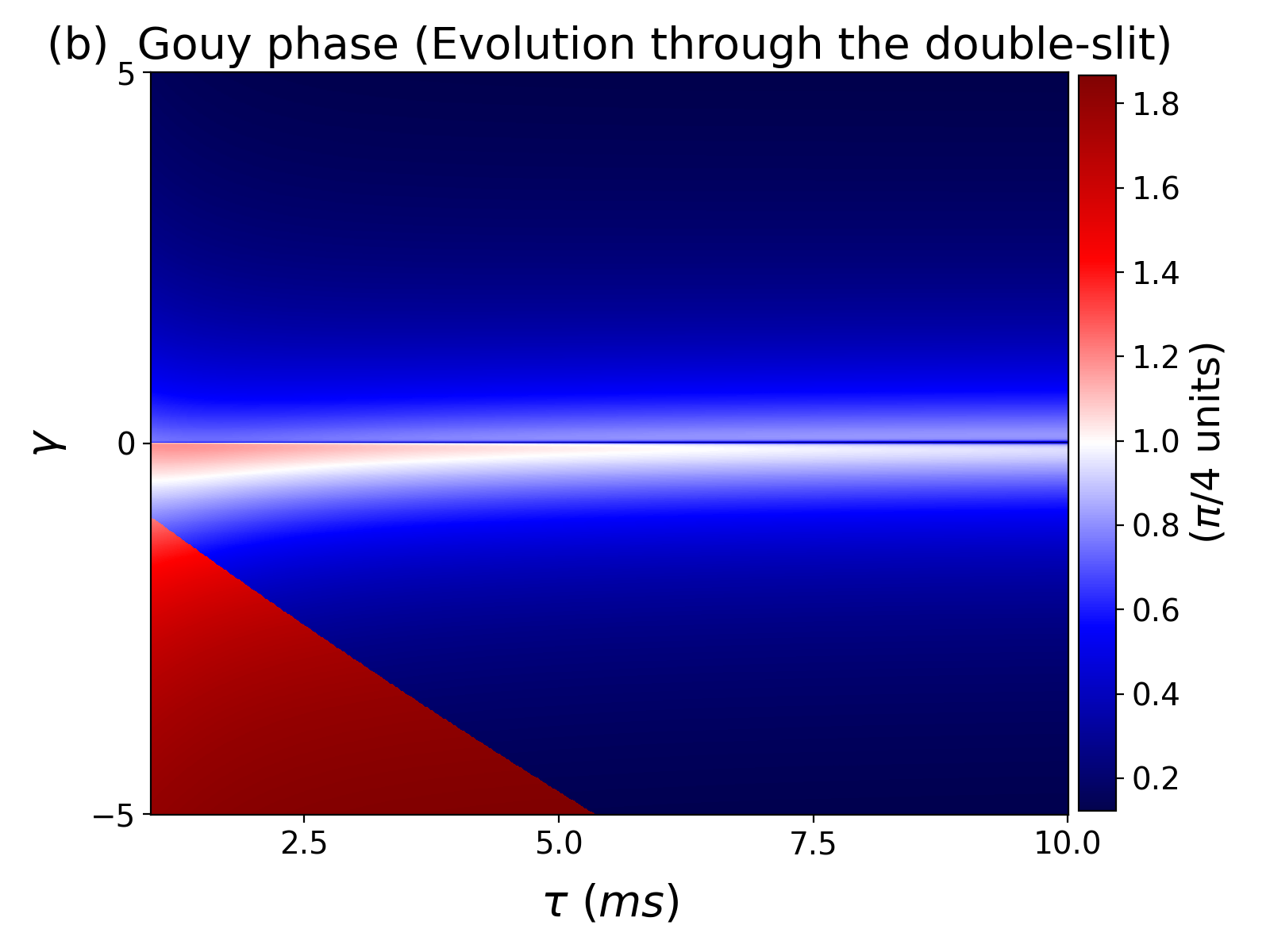}
\caption{(a) Absolute value of the Gouy phase difference, for the free evolution case, as a function of $\gamma$ and $t$. Note that, the phase effect is more apparent when the state is contractive $\gamma <0$, increasing further the transverse confinement of the wavepacket with respect to its propagation direction, and consequently, increasing the Gouy phase. (b) Absolute value of the Gouy phase difference, for
evolution through a double-slit, as a function $\gamma$ and $\tau$ for $t=50$ ms. The additional phase accumulation is influenced by the correlation parameter again, similar to the free-propagation scenario. For some values of $\gamma$, the phase has a greater magnitude, indicating specific settings where the phase effect is more evident.}\label{Figure1}
\end{figure}

 In Fig. \ref{Figure1} (a), we plot the absolute value of the Gouy phase difference, for the free evolution case, as a function of $\gamma$ and $t$. The behavior of the Gouy phase difference accumulated throughout the propagation period $t$ can be seen. Moreover, note the dependence with the correlation parameter $\gamma$, as expected, since it is responsible for modifying transversely the confinement condition of the wavepacket with respect to its propagation direction~\cite{LMarinhoPRA2020}, causing further phase accumulation for states with $\gamma \neq 0$. It should be noted that the phase effect is even more apparent for $\gamma <0$, a state known as contractive~\cite{YuenPRL1983} since it can be viewed as a squeezing agent in this case~\cite{LMarinhoPRA2020}. However, when $\gamma >0$, the result is the opposite.

We plot, in Fig. \ref{Figure1} (b), the cross-Wigner Gouy phase difference, for
evolution through a double-slit, as a function $\gamma$ and $\tau$ for $t=50$ ms . We observe the behavior of the Gouy phase difference acquired during the propagation time $\tau$. The additional phase accumulation is again dependent on the correlation parameter, just as in the free-propagation case. Notice that there are some values of $\gamma$ where the phase effect is more pronounced than others, indicating specific settings where the phase is more important.

\subsection*{Parameters of the cross-Wigner for free evolution}\label{AppendixA}

Here, we describe the parameters used to construct the real and imaginary parts of the cross-Wigner function while considering the free-evolution of a correlated Gaussian state (eq. (\ref{psi_free})) and the initial state (eq. (\ref{psi_0})), as shown in the main text.

\begin{gather}
A={\left(\frac{1}{2b^{2}}+\frac{1}{2\sigma_{0}^{2}}\right)^{2}+\left(\frac{m}{2\hbar r} -\frac{\gamma}{2\sigma_{0}^{2}}\right)^{2}}, \quad 
N^{-1}=\pi\sqrt{b\sigma_{0}}\sqrt[4]{\left(\frac{1}{2b^{2}}+\frac{1}{2\sigma_{0}^{2}}\right)^{2}+\left(\frac{m}{2\hbar r} -\frac{\gamma}{2\sigma_{0}^{2}}\right)^{2}},
\end{gather}

\begin{equation}
a_{1}=\left(\frac{1}{2b^{2}}+\frac{1}{2\sigma_{0}^{2}}\right)-a_{11},  \quad
a_{2}=-\left(\frac{m}{2\hbar r}-\frac{\gamma}{2\sigma_{0}^{2}}\right)+a_{22},  \quad
a_{3}=\frac{1}{A}\left(\frac{1}{2b^{2}}+\frac{1}{2\sigma_{0}^{2}}\right),  \quad
a_{4}=\frac{1}{A}\left(\frac{m}{2\hbar r}-\frac{\gamma}{2\sigma_{0}^{2}}\right),
\end{equation}

\begin{gather}
a_{5}=\frac{2}{A}\left(\frac{1}{2b^{2}}-\frac{1}{2\sigma_{0}^{2}}\right)\left(\frac{m}{2\hbar r}-\frac{\gamma}{2\sigma_{0}^{2}}\right) - \frac{2}{A}\left(\frac{m}{2\hbar r}+\frac{\gamma}{2\sigma_{0}^{2}}\right)\left(\frac{1}{2b^{2}}+\frac{1}{2\sigma_{0}^{2}}\right),    
\end{gather}

\begin{gather}
a_{6}=\frac{2}{A}\left(\frac{m}{2\hbar r}+\frac{\gamma}{2\sigma_{0}^{2}}\right)\left(\frac{m}{2\hbar r}-\frac{\gamma}{2\sigma_{0}^{2}}\right)+\frac{2}{A}\left(\frac{1}{2b^{2}}-\frac{1}{2\sigma_{0}^{2}}\right)\left(\frac{1}{2b^{2}}+\frac{1}{2\sigma_{0}^{2}}\right),    
\end{gather}

\begin{gather}
a_{11}=\frac{1}{A}\left[\left(\frac{1}{2b^{2}}-\frac{1}{2\sigma_{0}^{2}}\right)^{2}-\left(\frac{m}{2\hbar r}+\frac{\gamma}{2\sigma_{0}^{2}}\right)^{2}\right]   \left(\frac{1}{2b^{2}}+\frac{1}{2\sigma_{0}^{2}}\right) 
+\frac{2}{A}\left(\frac{1}{2b^{2}}-\frac{1}{2\sigma_{0}^{2}}\right)\left(\frac{m}{2\hbar r}+\frac{\gamma}{2\sigma_{0}^{2}}\right)\left(\frac{m}{2\hbar r}-\frac{\gamma}{2\sigma_{0}^{2}}\right),    
\end{gather}

\begin{gather}
a_{22}=-\frac{1}{A}\left[\left(\frac{1}{2b^{2}}-\frac{1}{2\sigma_{0}^{2}}\right)^{2}-\left(\frac{m}{2\hbar r}+\frac{\gamma}{2\sigma_{0}^{2}}\right)^{2}\right]  \left(\frac{m}{2\hbar r}+\frac{\gamma}{2\sigma_{0}^{2}}\right)\nonumber 
+ \frac{2}{A}\left(\frac{1}{2b^{2}}-\frac{1}{2\sigma_{0}^{2}}\right)\left(\frac{1}{2b^{2}}+\frac{1}{2\sigma_{0}^{2}}\right)\left(\frac{m}{2\hbar r}+\frac{\gamma}{2\sigma_{0}^{2}}\right).    
\end{gather}


\subsection*{Parameters of the cross-Wigner for evolution through a double-slit}\label{AppendixB}

In this part, we present the parameters that were utilized to describe the real and imaginary parts of the cross-Wigner function between the initial state (Eq. (\ref{psi_0})) and the superposition state at the detection screen (Eq. (\ref{superposition_state})).

\begin{equation}
B^2(t,\tau)= \frac{ \left(\frac{1}{\beta^2} + \frac{1}{b^2}
\right)^2 + \frac{m^2}{\hbar^2} \left(\frac{1}{ \tau} +
\frac{1}{r}\right)^2}{ (\frac{m}{\hbar \tau})^2
\left(\frac{1}{\beta^2} + \frac{1}{b^2}  \right)}, \quad 
R (t,\tau)= \tau \frac{ \left(\frac{1}{\beta^2} + \frac{1}{b^2}
\right)^2 + \frac{m^2}{\hbar^2} \left(\frac{1}{ \tau} +
\frac{1}{r}\right)^2}{ \frac{1}{\beta^4} + \frac{C}{\sigma^4_0 (t^2
+ \tau^2_0 + 2 \tau_0 t \gamma + t^2 \gamma^2)}}, \quad
\Delta(t,\tau)=\frac{\tau \sigma^2_0 d}{2 \tau_0 \beta^2 B^2},
\end{equation}
\begin{equation}
C=\left[ \tau^2_0 + \frac{t \tau^2_0}{\tau} + \tau^2_0 \gamma^2 +
\frac{\tau^3_0 \gamma}{\tau} + \frac{t \tau^2_0  \gamma^2}{\tau} +
\frac{2\tau^2_0 \sigma^2_0}{\beta} \right], \quad \theta(t,\tau)= \frac{m d^2 \left( \frac{1}{\tau} +
\frac{1}{r}\right)}{ 8 \hbar \beta^4 \left[ \left( \frac{1
}{\beta^2} + \frac{1}{b^2}\right)^2 + \frac{m^2}{\hbar^2} \left(
\frac{1}{\tau} + \frac{1}{r}\right)^2 \right ]},
\end{equation} 
\begin{equation}
D(t,\tau)=d \frac{\left(1+{\frac{\tau}{r}}\right)}{\left( 1+
\frac{\beta^2}{b^2}\right)}, \quad N^{\prime-1}=\pi\sqrt{B\sigma_{0}}\sqrt[4]{A^{\prime}} \sqrt{2 + 2 \exp \left[ -  \frac{D^2}{4 B^2}  - \Delta^2 B^2 \right]}
\end{equation}
\begin{gather}
A^{\prime}={\left(\frac{1}{2B^{2}}+\frac{1}{2\sigma_{0}^{2}}\right)^{2}+\left(\frac{m}{2\hbar R} -\frac{\gamma}{2\sigma_{0}^{2}}\right)^{2}},
\end{gather}
\begin{gather}
\alpha_1 = \frac{1}{A^{\prime}}\Bigg[\Bigg(\frac{1}{4B^4}-\frac{1}{4\sigma_0^4}\Bigg)+2\Bigg(\frac{m^2}{4\hbar^2 R^2} -\frac{\gamma^2}{4\sigma_0^4}\Bigg)\Bigg]  \Bigg(\frac{1}{2B^2}-\frac{1}{2\sigma_0^2}\Bigg)  - \frac{1}{A^{\prime}}\Bigg(\frac{m}{2\hbar R} +\frac{\gamma}{2\sigma_0^2}\Bigg)\Bigg(\frac{1}{2B^2}+\frac{1}{2\sigma_0^2}\Bigg), 
\end{gather}
\begin{gather}
\alpha_2 = -\frac{1}{A^{\prime}}\Bigg[\Bigg(\frac{1}{2B^2}-\frac{1}{2\sigma_0^2}\Bigg)^2-\Bigg(\frac{m}{2\hbar R} +\frac{\gamma}{2\sigma_0^2}\Bigg)^2\Bigg]  \Bigg(\frac{m}{2\hbar R} -\frac{\gamma}{2\sigma_0^2}\Bigg) + \frac{2}{A^{\prime}}\Bigg(\frac{m}{2\hbar R} +\frac{\gamma}{2\sigma_0^2}\Bigg)\Bigg(\frac{1}{4B^4}-\frac{1}{4\sigma_0^4}\Bigg), 
\end{gather}
\begin{gather}
b_1 = \frac{1}{A^{\prime}}\Bigg(\frac{D^2}{16B^4}-\frac{\Delta^2}{4}\Bigg)\Bigg(\frac{1}{2B^2}+\frac{1}{2\sigma_0^2}\Bigg) + \Bigg(\frac{m}{2\hbar R}-\frac{\gamma}{2\sigma_0^2}\Bigg)\Bigg(\frac{\Delta D}{4B^2A^{\prime}}\Bigg), \quad
b_2 = \frac{1}{A^{\prime}}\Bigg(\frac{1}{2B^2}+\frac{1}{2\sigma_0^2}\Bigg), \quad 
b_3 = \frac{1}{2B^2}+\frac{1}{2\sigma_0^2}-\alpha_1,
\end{gather}
\begin{equation}
b_4 = \frac{1}{A^{\prime}}\Bigg(\frac{m}{\hbar R\sigma_0^2}+\frac{\gamma}{\sigma_0^2B^2}\Bigg), \quad 
b_5 = \Bigg[\Bigg(\frac{1}{4B^4}-\frac{1}{4\sigma_0^4}\Bigg)+\Bigg(\frac{m^2}{4\hbar^2 R^2} -\frac{\gamma^2}{4\sigma_0^4}\Bigg)\Bigg]\Bigg(\frac{D}{2B^2A^{\prime}}   \Bigg) -\frac{\Delta}{2A^{\prime}}\Bigg(\frac{m}{\hbar R\sigma_0^2} +\frac{\gamma}{\sigma_0^2B^2}\Bigg)              -\frac{D}{2B^2}, 
\end{equation}
\begin{gather}
b_6 = \Bigg(\frac{m}{2\hbar R}-\frac{\gamma}{2\sigma_0^2}\Bigg)\Bigg(\frac{D}{2B^2A^{\prime}}\Bigg)-\Bigg(\frac{1}{2B^2}+\frac{1}{2\sigma_0^2}\Bigg)\Bigg(\frac{\Delta}{A^{\prime}}\Bigg), \quad
b_7=\alpha_2 - \frac{m}{2\hbar R}+\frac{\gamma}{2\sigma_0^2},
\end{gather}	
\begin{gather}
b_8 =\frac{1}{A^{\prime}}\Bigg[\Bigg(\frac{1}{2B^4}-\frac{1}{2\sigma_0^4}\Bigg) +\Bigg(\frac{m^2}{2\hbar^2 R^2}-\frac{\gamma^2}{2\sigma_0^4}\Bigg)\Bigg], \quad
b_9 = \frac{1}{A^{\prime}}\Bigg(-\frac{D^2}{16B^4}+\frac{\Delta^2}{4}\Bigg)\Bigg(\frac{m}{2\hbar R}-\frac{\gamma}{2\sigma_0^2}\Bigg) + \Bigg(\frac{1}{2B^2}+\frac{1}{2\sigma_0^2}\Bigg)\Bigg(\frac{\Delta D}{4B^2A^{\prime}}\Bigg),
\end{gather}	
\begin{gather}
b_{10} = \frac{1}{A^{\prime}}\Bigg(\frac{m}{2\hbar R}-\frac{\gamma}{2\sigma_0^2}\Bigg), \quad
b_{11} = \frac{\Delta}{2A^{\prime}}\Bigg[ \Bigg(\frac{1}{2B^4}-\frac{1}{2\sigma_0^4} \Bigg)  + \Bigg(\frac{m^2}{4\hbar^2 R^2}-\frac{\gamma^2}{2\sigma_0^4}\Bigg) \Bigg]  +  \Bigg(\frac{m}{2\hbar R\sigma_0^2}+\frac{\gamma}{2\sigma_0^2B^2}\Bigg)\Bigg(\frac{D}{2B^2A^{\prime}} \Bigg) - \Delta,
\end{gather}
and
\begin{gather}
b_{12} = \Bigg(\frac{1}{2B^2}+\frac{1}{2\sigma_0^2} \Bigg)\Bigg(\frac{D}{2B^2A^{\prime}}\Bigg) + \Bigg(\frac{m}{2\hbar R}-\frac{\gamma}{2\sigma_0^2}\Bigg)\Bigg(\frac{\Delta}{A^{\prime}}\Bigg).
\end{gather}

\bibliography{citations}

\section*{Acknowledgements}
L.S.M. acknowledges Funda\c{c}\~ao de Amparo \`a Ci\^encia e Tecnologia do Estado de Pernambuco and Office of Naval Research (N62909-23-1-2014). P.R.D. acknowledges support from the Foundation for Polish Science (IRAP project, ICTQT, Contract No. MAB/2018/5, co-financed by EU within Smart Growth Operational Programme).  C.H.S.V. acknowledges CAPES (Brazil) and Federal University of ABC for the financial support. I.G.P. acknowledges Grant No. 306528/[2023-1] from CNPq.

\section*{Author contributions}

L.S.M., P.R.D., C.H.S.V. and I.G.P. contributed equally for the calculations and for writing the manuscript.  All authors read and approved the final manuscript.

\section*{Competing interests}
Te authors declare no competing interests.

\section*{Additional information}
Correspondence and requests for materials should be addressed to L.S.M

\section*{Data availability}

The datasets used and/or analysed during the current study available from the corresponding author on reasonable request.

\end{document}